# Integrated vortex-assisted electroporation platform with enhanced throughput for genetic delivery to primary cells

Hyun Woo Sung[1] and Soojung Claire Hur[2-4]

[1]*Department of Chemical and Biomolecular Engineering, Johns Hopkins University, Baltimore, Maryland, United States of America*

[2]*Department of Mechanical Engineering, Johns Hopkins University, Baltimore, Maryland, United States of America*

[3]*The Sidney Kimmel Comprehensive Cancer Center, Johns Hopkins Hospital, Baltimore, Maryland, United States of America*

[4]*Institute of NanoBioTechnology, Johns Hopkins University, Baltimore, Maryland, United States of America*

**Abstract**

Primary human cells offer the most faithful representation of native human physiology, yet their practical utility is constrained by the difficulty of introducing exogenous genetic material. Electroporation provides a promising non-viral gene delivery approach; however, conventional bulk systems lack the uniformity and integration required for heterogeneous primary cell samples. Here, we present a vortex-assisted electroporation platform integrating size-selective cell trapping with enhanced throughput, parameter optimization across buffer and electrical conditions, and robust delivery of plasmid DNA and in vitro-transcribed mRNA in primary human cells. This integrated platform provides a unified workflow that addresses sample heterogeneity, throughput demands, and delivery efficiency, enabling broader implementation of non-viral gene delivery into primary cells for research and translational applications.

## 1 Introduction

### 1.1 Motivation

Primary human cells, directly isolated from human tissues or organs, most faithfully recapitulate native human physiology and are therefore essential for studying patient-oriented disease mechanisms. Their broader utility, however, is constrained by challenges in introducing exogenous genetic material for genetic manipulation and functional studies. Compared to immortalized cell lines, primary cells exhibit limited proliferative capacity[1,2] and substantial heterogeneity in membrane properties and cell size[3], complicating efficient intracellular delivery.

Chemical and viral gene delivery methods introduce well-known limitations, including reliance on exogenous carrier materials, cytotoxicity, immunogenicity, potential genome integration, and complex pre-selection workflows[4]. Electroporation represents a promising non-viral and carrier-free physical approach that supports delivery of diverse cargoes, including plasmids, mRNA, and proteins[5] with performance tunable through electric field strength, pulse duration, and buffer composition[6–10]. However, conventional bulk electroporation systems require high voltages, generate non-uniform electric fields, and lack the precision needed to accommodate heterogeneous primary cell samples, resulting in inefficient and irreproducible delivery[5,11].

Microscale electroporation integrated within microfluidic platforms has addressed many of these limitations by enabling precise control of electric field generation and localization, fluid handling, and cell manipulation[12–15]. Several microfluidic systems have demonstrated successful gene delivery to primary cells by employing designs such as microelectrode arrays[16], continuous flow-through electroporation systems[14,15,17,18], and droplet-based formats[19,20]. Nevertheless, most existing platforms focus exclusively on the transfection step and lack upstream integration of size-selective cell isolation or enrichment, effectively assuming that input samples are already pre-purified and tightly size-distributed. As a result, their utility for processing heterogeneous, real-world biological samples is limited. This lack of workflow integration represents a significant barrier for applications involving heterogeneous samples, where throughput and processing must be accommodated within a unified system.

Building on this need, we previously developed a platform that combines vortex-based cell trapping with tunable electroporation[21–24]. Vortex trapping has been used to enrich rare human cells from complex biofluids[25–27], and its integration with electroporation has demonstrated feasibility for multi-molecular delivery[21–24]. However, earlier demonstrations primarily established the trapping-transfection integration itself[21–23], whereas more recent work extended this approach to immortalized cells spiked into whole blood, capturing key aspects of heterogeneous biological samples[24].

Here, we present a fully integrated, enhanced-throughput vortex-assisted electroporation platform (**Figure 1**) designed for size-selective primary cell transfection. Building on a commercially validated vortex trapping geometry demonstrated for rare-cell enrichment in complex biofluids[25–27], this platform incorporates a redesigned microscale electrode array that substantially increases processing throughput while maintaining precise control over electric parameters. Key electroporation variables, including buffer composition, electric field strength, and cargo concentration, are systematically examined using human primary cells. Platform versatility is further explored through delivery of both plasmid DNA and in vitro-transcribed (IVT) mRNA, enabling assessment of compatibility across genetic cargos of differing size and expression characteristics.

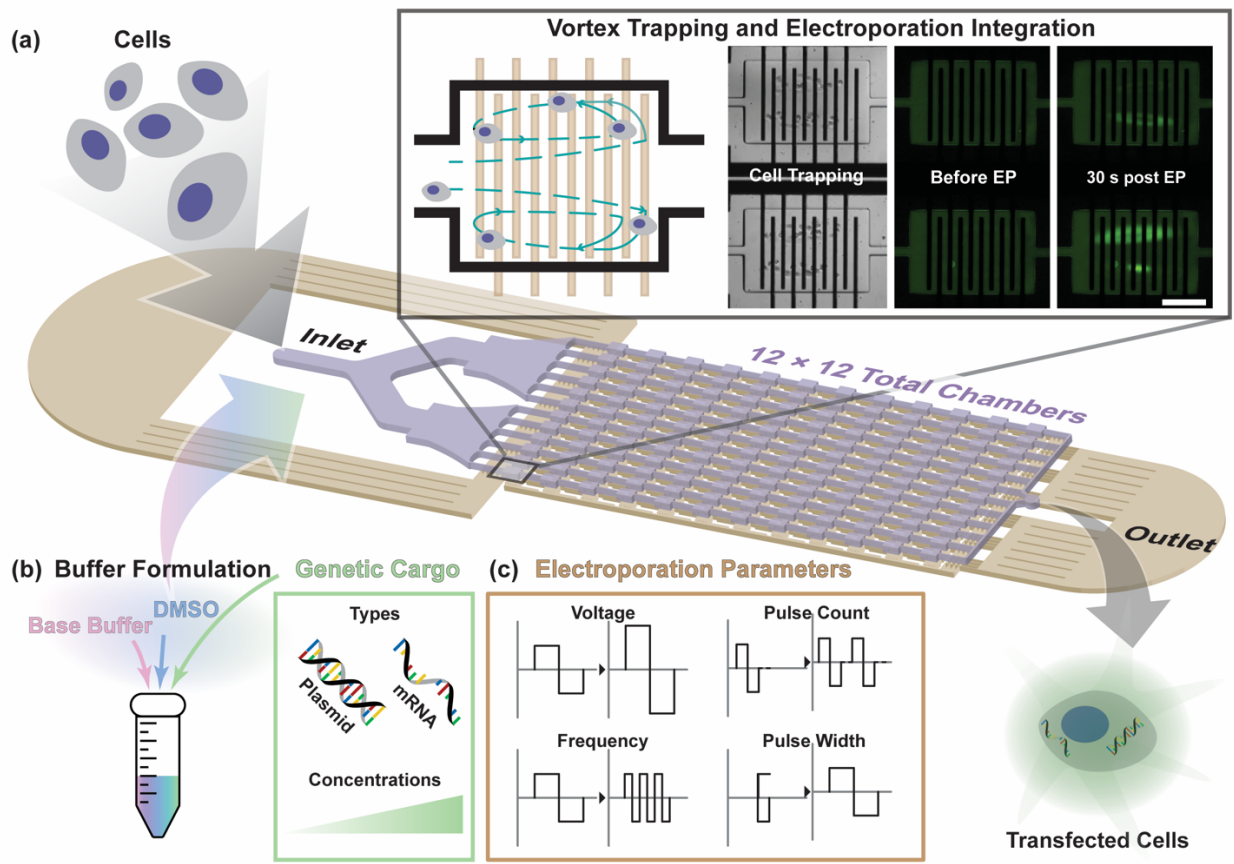

**Figure 1: System overview. (a)** Schematic of the vortex-assisted electroporation platform with enhanced throughput. Cells enter through the inlet and are hydrodynamically trapped within microscale vortex chambers in a size-selective manner, retaining cells above a designed diameter threshold for electroporation using an integrated electrode array. Representative brightfield and fluorescence images illustrate cell trapping and electroporation before and after pulse application. Following electroporation, transfected cells are collected at the outlet. **(b)** Buffer formulation and cargo modularity. The electroporation buffer is assembled from commonly used laboratory reagents, including a base buffer supplemented with DMSO and genetic cargo. This formulation strategy provides flexibility in selecting and adjusting buffer compositions across different types, while supporting delivery of diverse cargo types, including plasmid DNA and mRNA, and enabling performance tuning through cargo concentration. **(c)** Electroporation parameter control. Voltage amplitude, pulse count, pulse frequency, and pulse width are independently tunable within the platform, enabling precise control of electrical stimulation during electroporation.

## 2    Results

### 2.1    Balancing Throughput and Field Uniformity in Electroporation Array Design

Enabling enhanced-throughput electroporation while preserving field precision and consistency required a new electrode array architecture that extends the prior 4×10 layout (4 parallel transverse rows, 10 longitudinal columns; 40 chambers)[23,24]. The updated array expands processing capacity while addressing critical engineering constraints, including chamber-to-chamber electric field uniformity and voltage drop across the inlet and outlet routing regions, which can otherwise compromise field uniformity in parallelized electroporation systems. A primary goal of the redesign was ensuring that the applied voltage translated efficiently and uniformly into consistent electric fields across all electroporation chambers — an essential requirement for reproducible gene delivery in parallelized systems.

Design efforts were shaped by multiple physical and fabrication constraints: (1) the maximum device footprint permitted by standard microscopic slides, (2) feature size selections made to ensure robust, high-yield photolithographic fabrication, (3) chamber spacing requirements necessary for fully developed vortex flow[25], and (4) electrode trace width and thickness requirements that preserve exposed glass for leak-proof PDMS-to-glass bonding. These constraints restricted the allowable array dimensions and demanded careful optimization of the routing pathways within confined regions.

Several candidate electrode array configurations were evaluated to balance cell-trapping capacity and electric field uniformity. Each layout contained three main functional sections: an inlet routing pathway, the electrode array, and an outlet routing pathway (**Figure 2a**). Three array configurations – 16×12 (VTX-1 style, 192 chambers)[25], 16×9 (144 chambers), and the final 12×12 (144 chambers) – were compared to assess the tradeoffs between field uniformity and electrical efficiency. Each chamber incorporated five pairs of interdigitated electrodes that generated localized electric fields for electroporating vortex-trapped cells (**Figure 2b**). Informed by prior work[23], inlet and outlet electrode sections connecting to anode and cathode pads were modified for each geometry to maintain electrical performance while accommodating layout-dependent routing.

The overall electrode schematics were converted into equivalent Simulation Program with Integrated Circuit Emphasis (SPICE) models representing resistor networks in series and parallel (**Figure 2c**). Uniform and efficient electroporation requires both minimal chamber-to-chamber voltage variability and high voltage efficiency, defined as the proportion of input voltage dropped across the array. These metrics were quantified for each configuration, and the input voltage needed to achieve a mean electric field of 900V/cm

(typical for reversible electroporation of mammalian cells[6,28,29]) was determined by SPICE circuit simulation.

Prior experience[23] indicated that minimizing chamber-to-chamber variability benefits from bifurcated inlet and outlet routing pathways that equalize resistance across the array (**Figure 2d**). This approach was first applied to the VTX-1-style 16×12 chamber layout[25]. Although this geometry maximized cell trapping throughput, SPICE simulations revealed substantial resistive losses along extended routing lines created by 12 electrodes in series. Generating a 900 V/cm mean electric field required 39.1 V (voltage efficiency of 48.8 ± 0.87%) and produced 7.70% voltage variability, indicating inefficient and inconsistent voltage transfer (**Figure 2f**).

Shortening the series path yielded the 16×9 configuration, which reduced the required input voltage to 32.2 V and correspondingly increased voltage efficiency to 53.4 ± 0.80%, while lowering chamber-to-chamber variability to 4.51% (**Figure 2f**). However, even the reduced 32.2 V required for the 16×9 layout still exceeded the structural tolerance of the micropatterned electrode traces. Elevated local current densities induce electrochemical erosion, producing voids and ultimately compromising device stability[30,31] (Figure S1).

An alternative 12×12 configuration incorporating shorter routing paths and widened inlet/outlet traces was evaluated (**Figure 2e**). This layout reduced total routing-path resistance by 2.5-fold relative to the 16×9 geometry, allowing a greater fraction of input voltage to drop across the array. Achieving a 900 V/cm field required only 24.0 V (**Figure 2f**) and voltage efficiency increased to 79.4 ± 2.28%. Although chamber-to-chamber variability rose slightly to 11.1%, the reduction in required input voltage mitigated voltage-dependent resistive losses and improved device stability.

Because this redesign altered the array geometry and routing architecture, its impact on throughput and trapping performance was next evaluated. Benchmarking against the commercially validated VTX-1 (16×12) configuration[25,26] enabled direct comparison of throughput under electroporation-integrated conditions, where additional electrical and material constraints limit achievable flow rates. The VTX-1 system reached flow rates of 8 mL/min ($Re$ = 170.6) using a multi-syringe pump, whereas the electroporation-integrated 12×12 device was constrained to 5.2 mL/min ($Re$ = 140.1) due to PDMS delamination at the micropatterned Au-glass interface. As expected from the reduced chamber count (192 to 144), the overall trapping efficiency of the 12×12 device was lower than that of VTX-1 (**Figure 2g**). When normalized by chamber number, per-chamber trapping remained largely comparable (0.11% vs. 0.14% captured per chamber), indicating preservation of size-selective trapping behavior in the 12×12 layout (Figure S2). This maintained trapping performance, together with the substantially improved electrical efficiency and device stability, supported selection of the 12×12 layout architecture for downstream electroporation experiments.

Overall, the 12×12 configuration provided the most favorable balance between electric field generation, resistive performance, and operational robustness. Final schematic and resistive networks for the 12×12 array are provided in Figure S3, and this configuration was selected for all downstream electroporation experiments.

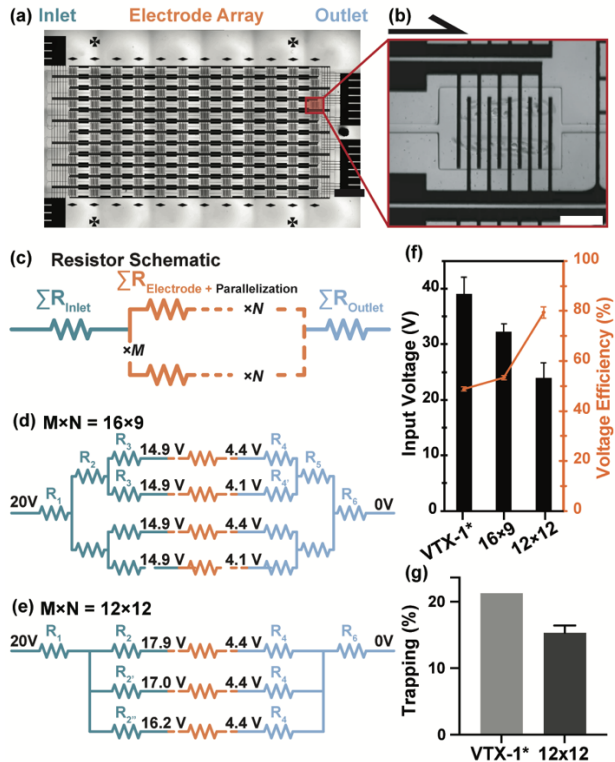

**Figure 2. Design and electrical optimization of an enhanced-throughput electroporation array. (a)** Brightfield image of the electrode layout on the electroporation chip, consisting of an inlet (teal), an electroporation array (orange), and outlet routing section (light blue). Scale bar = 2 mm. **(b)** Magnified brightfield image of a single electroporation chamber, showing cells trapped in vortices circulating around interdigitated electrodes. Scale bar = 250 μm. **(c)** Simplified resistance network representation of the electrode array, partitioned into inlet routing, electrode array, and outlet routing regions. Array parallelization (M×N) denotes the number of electroporation chambers generated through lateral (M) and longitudinal (N) replication of a base electrode unit. **(d)** Resistance network corresponding to the 16×9 layout, in which symmetric bifurcation of the inlet routing electrode produces two dominant resistance paths, reducing chamber-to-chamber variability. **(e)** Resistance network corresponding to the 12×12 layout, incorporating shortened routing paths and reduced inlet and outlet

resistance, thereby shifting a greater fraction of the applied voltage drop onto the electrode array and improving voltage efficiency at the expense of increased resistive asymmetry. For all resistance networks shown, annotated voltage values (black) indicate simulated voltage drops between regions under a 20 V applied potential. Only the upper half of the network (M/2) is illustrated, as symmetry yields identical resistive behavior in both halves. **(f)** Input voltage required to achieve a mean electric field of 900 V/cm within the electrode array (left axis) and corresponding voltage efficiency (right axis). The 12×12 configuration requires substantially lower input voltage than the 16×9 and VTX-1 configurations, while exhibiting higher voltage efficiency. Error bars for input voltage indicate spatial voltage variability across chambers (standard deviation) whereas voltage efficiency error bars represent row-to-row variability. **(g)** First-cycle cell trapping efficiency comparison between the ultra-high throughput device (VTX-1; data referenced from refs.[25,26], purification-only platform) and the integrated 12×12 electroporation device. Error bars represent mean ± SEM (n = 3).

## 2.2 Integrated Trapping-Electroporation Performance Characterization

The feasibility of vortex-assisted electroporation has been demonstrated previously, including with our earlier 4x10 platform, which enabled efficient multi-cargo delivery through size-selective pre-concentrating and real-time control of electrical parameters[21,23,32]. Building on this foundation, the redesigned 12x12 array evaluated using MCF7 cells, a model cell line previously used to assess both vortex trapping and microscale electroporation[23,25], to determine whether integrated trapping and electroporation performance is preserved under enhanced-throughput operation.

Electroporation efficiency and viability quantified using YOYO-1 and Calcein AM Red revealed the expected trade-off between membrane permeabilization and cell integrity: increasing input voltage enhanced delivery efficiency but also intensified membrane damage and electrochemical stress, resulting in reduced viability and overall cell recovery (**Figure 3a**). The highest electroporation efficiency observed was 85.9% at 23 V, accompanied by reduced viability (41.5%) and recovery (78.0%), respectively. Consistent with our prior experience using the lower-throughput 4x10 vortex-assisted electroporator[23,32], a modest reduction in transfection efficiency was necessary to achieve practical levels of viability and recovery. Accordingly, 20 V provided the most balanced operating condition, yielding 74.9% delivery efficiency, 65.7% viability, and 88.0% recovery (**Figure 3a**, blue shading). These results illustrate the characteristic tradeoff between delivery efficiency and cell health in electroporation systems.

Parallelization of the electrode array increased sample-processing capacity while preserving the Reynolds number required for stable vortex trapping in each row. Under these conditions, integration of the solution-exchange system enabled sustained operation at 5.2 mL/min, representing a five-fold increase over the 4×10 platform's 1.0 mL/min throughput (**Figure 3b**). Despite this increase in processing capacity, the 12x12 array maintained electroporation performance comparable to the 4x10 system across delivery efficiency and viability metrics (**Figure 3c-e**), demonstrating that parallelization enhances throughput without substantially compromising delivery outcomes. These observations indicate that the redesigned 12x12 array preserve effective vortex trapping and electroporation performance under enhanced-throughput operation, providing a baseline for subsequent studies in heterogeneous primary cell samples.

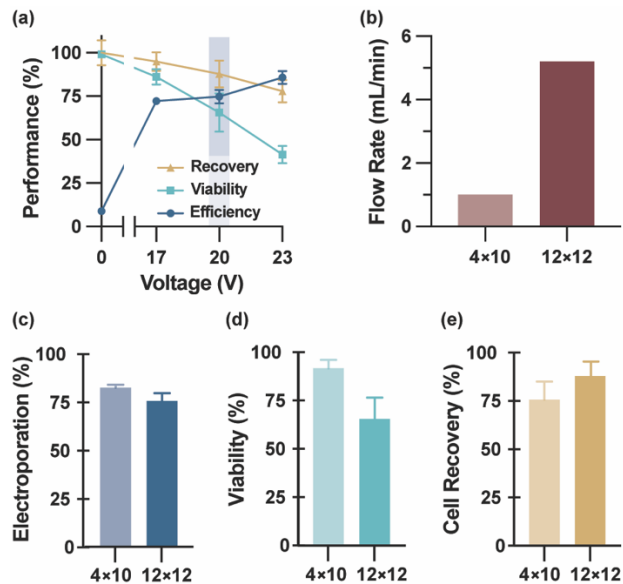

**Figure 3. Electroporation performance and throughput scaling**. **(a)** Electroporation performance of MCF-7 cells quantified by membrane-impermeable dye uptake with post-electroporation viability counterstaining. The gating strategy for identifying successful electroporated cells is shown in Figure S4. Nuclear labeling performed prior to processing enabled accurate cell enumeration. The shaded gray region denotes the optimal operating voltage range. Data are shown as mean ± SEM (n = 3; >300 cells per replicate). **(b)** Sample throughput comparison between 4×10 and 12×12 chamber configurations operated at 40 psi, demonstrating a >5-fold increase in throughput with array scaling. Quantitative comparison of **(c)** electroporation efficiency, **(d)** cell viability, and **(e)** cell recovery between 4×10 and 12×12 devices under optimized conditions. Data are presented as mean ± SEM (n = 3).

### 2.3 Optimization of Electroporation Parameters for Primary Human Cells

Vortex-assisted electroporation has demonstrated robust performance in immortalized cell lines[21,23,32], but extension to human primary cells requires additional optimization given their fragile membranes, limited proliferative capacity, and heightened sensitivity to stress[3]. We examined whether electrical and chemical parameter tuning could enable effective transfection in primary cells, using human mammary fibroblasts (HMFs) to define operating conditions that support membrane permeabilization.

Quantification of electroporation efficiency, viability, and cell recovery using membrane-impermeable YOYO-1 uptake (**Figure 4a**) revealed trends consistent with immortalized cell lines: increasing input voltage enhanced membrane permeabilization and delivery efficiency but concurrently produced voltage-dependent reductions in viability and recovery. A balanced operating regime emerged at intermediate voltages (16-20V), where delivery efficiency increased substantially while maintaining acceptable levels of cell health, establishing electrical conditions for subsequent plasmid delivery studies.

Optimization efforts were then extended to electrical parameters using Dulbecco's Phosphate Buffered Saline (DPBS) as the electroporation buffer with 50 μg/mL ZsGreen plasmid. While DPBS supported ZsGreen delivery into HEK 293 cells with 37.5% efficiency (normalized efficiency 0.617; Figure S6), primary human mammary fibroblasts (HMFs) exhibited minimal transfection (0.25%; normalized efficiency of 0.019, Figure S6), even after extensive tuning of voltage, pulse width, pulse counts, and waveform (Figure S7). These results indicate that electrical-parameter optimization alone was insufficient to enable effective transfection in primary cells, motivating investigation of chemical parameter modulation in addition to electrical tuning.

Electroporation buffer composition is known to strongly influence membrane permeabilization and cell survival[7,33,34]. Conventional bulk electroporation systems often rely on proprietary buffers to mitigate cellular stress caused by high-voltage pulses. We therefore examined non-proprietary buffer formulations for primary cell transfection, beginning with dimethyl sulfoxide (DMSO), a membrane-modifying agent[35] known to stabilize hydrophilic pores and lower the electric field threshold for permeabilization[36–38]. However, supplementing DPBS with 1% (v/v) DMSO produced negligible improvement in HMF transfection (Figure S8a), indicating that DPBS-based formulations were inadequate even in the presence of a membrane-permeabilizing additive.

Given the limitation of DPBS, Opti-MEM™, a reduced-serum, sodium bicarbonate-buffered medium reported to support cell viability during electroporation[9,39], was next evaluated as an alternative base buffer. Direct comparison of Opti-MEM + DMSO versus DPBS + DMSO revealed a pronounced voltage-dependent enhancement in transfection efficiency, with

Opti-MEM providing the greatest improvement at higher voltage and reduced gains at lower voltages (**Figure 4b**). To facilitate comparison across conditions, all results were normalized to our baseline of 1% (v/v) DMSO in DPBS at each formulation's optimal voltage. Using the DMSO-Opti-MEM buffer, we then systematically assessed the combined effects of voltage and plasmid concentration on HMF transfection.

Guided by the YOYO-1 permeabilization data indicating comparable membrane accessibility at intermediate voltages, conditions were grouped into low (<18V) and high (≥18V) voltage regimes. Increasing plasmid concentration consistently enhanced delivery, with the highest efficiency achieved at 200 μg/mL plasmid under high-voltage conditions, corresponding to a 46.5-fold improvement over baseline and reaching 88% of Lipofectamine performance (Figure S3.3.4a). However, higher voltages accompanied by reduced downstream cell recovery (Figure S3.3.4b), highlighting the inherent tradeoff between maximizing delivery efficiency and preserving viable primary cells. Within this framework, a plasmid concentration of 200 μg/mL combined with intermediate applied voltage (16 -18 V) provided the most balanced performance for HMF transfection, excluding 20V due to reduced recovery.

Efficient primary cell electroporation requires coordinated optimization of electrical and chemical parameters. Although the DMSO-Opti-MEM formulation substantially improved HMF delivery, transfection requirements vary widely across primary cell types. The platform's multi-inlet architecture enables real-time formulation mixing and precise control of solution composition[21,32], positioning the system for future automated and combinatorial screening of electroporation conditions.

## 2.4 Cargo and Cellular Factors Governing Primary Cell Transfection

Robust electroporation performance across diverse applications requires understanding how both cargo characteristics and cellular state influence delivery outcome. Building on the reporter-based validation above, we next extended our analysis to a large, non-reporter plasmid cargo: a hemagglutinin (HA)-tagged human telomerase reverse transcriptase (hTERT) construct (9.0 kb) delivered into HMF cells. hTERT, the catalytic subunit of the telomerase complex, is tightly regulated in primary somatic cells[40–42]. Given that cellular aging alters fibroblast membrane properties and gene expression[43–45], electroporation outcomes were compared between low-passage (population doubling (PD) < 14) and high-passage (PD 14–30) HMFs, which differ in proliferative activity as confirmed by Ki67 staining (Figure S9).

Transfection efficiencies of 4.07% (low-passage) and 7.84% (high-passage) were observed at 16 V (E = 960 V/cm), while increasing the voltage to 18 V (E = 1080 V/cm) reduced expression to 2.15% and 3.82%, respectively, indicating non-monotonic voltage dependence and diminished delivery performance at higher field strengths (**Figure 4c-d**). Higher-passage HMF cells consistently exhibited greater hTERT expression, suggesting that senescence-associated changes may enhance susceptibility to electroporation. Overall transfection efficiencies were lower than those observed with ZsGreen, likely reflecting the larger size of the hTERT plasmid (9.0 kb vs. 4.7 kb). This size-dependent decrease is consistent with prior reports showing reduced delivery efficiency for larger plasmids (6–16 kb) due to increased membrane disruption requirements and slower post-electroporation recovery[46]. These results demonstrate that the platform enables delivery and expression of large, non-reporter genetic cargo, and that delivery efficiency depends jointly on plasmid size and cellular state—important considerations for tailoring electroporation protocols to diverse primary cell types.

To assess platform versatility beyond plasmid DNA, we examined mRNA electroporation, which bypasses nuclear import and transcription and enables rapid protein expression[47]. As chemically modified mRNA can improve translation efficiency and reduce innate immune activation[48,49], we synthesized three distinct eGFP mRNA constructs with different cap and nucleoside chemistries and identified an N1-methylpseudouridine (N1MePsU)-substituted, CleanCap AG-capped variant as the highest-performing candidate (Figure S10). Electroporation of this optimized construct across mRNA concentrations ranging from 0-30 µg/mL yielded a maximum expression efficiency of 76.2% at 18V (E=1080 V/cm), corresponding to 78% of the efficiency achieved using Lipofectamine (**Figure 4e-f**). Lower mRNA doses further reduce cytotoxicity, highlighting an advantage of mRNA over plasmid DNA for primary cell applications.

These results indicate that the lower delivery efficiencies observed for plasmid cargos (ZsGreen and hTERT), relative to mRNA, arise from cargo-specific limitations—such as size, nuclear import, and transcriptional dependence—rather than constraints of electroporation platform itself. In contrast, mRNA (~900 nt) bypasses these barriers, enabling rapid and efficient protein expression with reduced cytotoxic burden. The platform supports delivery of diverse nucleic acid cargos with efficiencies approaching chemical transfection methods while operating at five-fold higher throughput than prior vortex-assisted electroporation systems. This performance positions the platform for multimodal genetic delivery applications in primary cells.

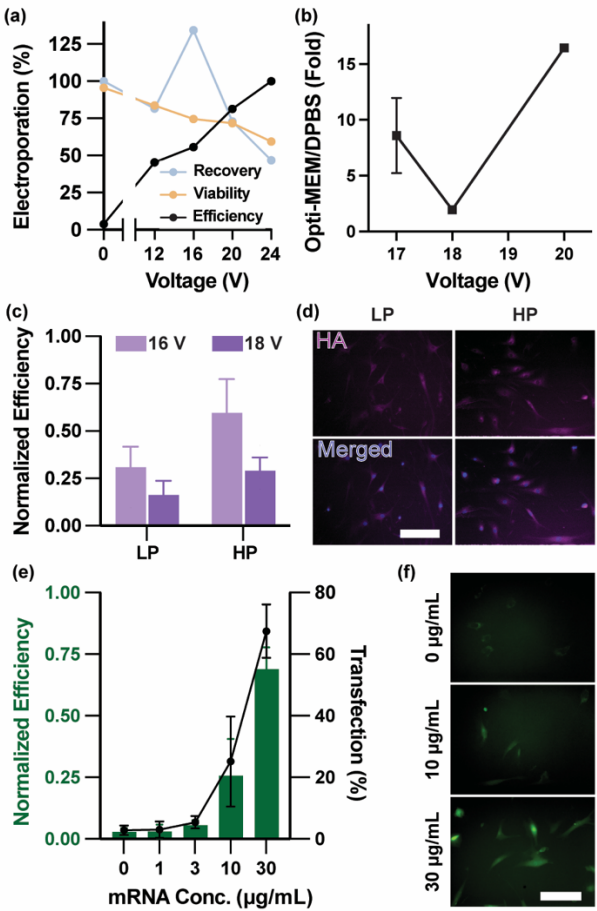

**Figure 4. Electrical and Chemical Optimization Enables Multimodal Gene Delivery in Primary Cells**. **(a)** Electroporation performance of human mammary fibroblasts (HMFs) in DPBS as a function of applied voltage, quantified by delivery efficiency, viability and recovery using membrane-impermeable YOYO-1 uptake. **(b)** Fold change in 50 µg/mL ZsGreen plasmid transfection efficiency using Opti-MEM + DMSO relative to DPBS + DMSO across voltages, a voltage-dependent enhancement, with Opti-MEM + DMSO consistently outperforming DPBS + DMSO across the tested voltage range. **(c)** hTERT plasmid

transfection efficiency in low-passage (LP, PD<14) and high-passage (HP, PD>14) HMFs at 16V and 18V, normalized relative to Lipofectamine-mediated delivery. Cells were electroporated in the Opti-MEM + DMSO with 200 µg/mL plasmid and immunostained 48 hr post-electroporation. **(d)** Immunofluorescence imaging of HA-tagged hTERT expression in LP and HP HMFs, showing DAPI nuclear staining, HA immunostaining, and merged channels. Scale bars = 100 µm. **(e)** Dose-response analysis of eGFP mRNA electroporation in HMFs. The left y-axis shows normalized efficiency relative to an mRNA Lipofectamine control (Messenger Max, 0.1 µg/well), and the right y-axis shows absolute transfection efficiency. **(f)** Representative fluorescence microscopy images showing dose-dependent eGFP expression following mRNA electroporation (0, 10, 30 µg/mL). Scale bar = 100 µm.

## 3 Conclusions

In this study, we developed an enhanced-throughput vortex-integrated electroporation platform for primary cell transfection by integrating size-selective cell trapping, on-chip solution exchange, and systematic optimization of electrical and chemical parameters. Redesigning the electrode array architecture enabled a five-fold increase in processing throughput, while preserving real-time tunability, carrier-free operation, and compatibility with sensitive cell types. The selected 12x12 array improved voltage efficiency and maintained stable performance at reduced input voltage.

Sequential optimization of electroporation parameters revealed that both buffer composition and cargo-specific properties critically govern delivery efficiency. An optimized Opti-MEM/DMSO formulation enabled efficient plasmid transfection in human mammary fibroblasts, and further tuning yielded efficiency up to 88% and 78% of Lipofectamine benchmarks for DNA and mRNA delivery, respectively, which generally rely on pre-selected, homogeneous cell populations. The platform successfully delivered both reporter and functional genes, including large plasmids such as hTERT, and uncovered passage-dependent differences in primary cell transfection. Complementary mRNA studies highlighted the role of cargo chemistry, molecular size, and electric field strength in dictating efficiency-viability tradeoffs.

Several aspects of this work present opportunities for further refinement. Validation was limited to a defined set of cell types, the current assessment emphasized short-term expression rather than long-term functionality, and throughput remains bounded by material constraints and modest chamber-to-chamber variability. Ongoing and future efforts will extend evaluation across additional primary cells, incorporate functional cargos, and leverage the multi-inlet architecture for *in situ* combinatorial buffer screening. Parallel advances in materials and system design will target improved stability and increased parallelization, while longitudinal assays will assess phenotypic durability.

Our work establishes a versatile and scalable electroporation platform for primary cells, enabling efficient delivery of diverse nucleic acid cargos and real-time optimization of delivery conditions. The integrated fluidic architecture provides a foundation for automated, multimodal genetic engineering workflows.

## 4 Methods

### 4.1 Device Fabrication

The device consists of a polydimethylsiloxane (PDMS) layer with the vortex cell-trapping chambers[23,25] enclosed by a glass slide with a microfabricated gold electrode array patterned on its surface. Detailed fabrication procedures are described in our previous report[24]. In brief, the device consists of a PDMS layer (cat#4019862, Dow) plasma-bonded to a micropatterned Au-electrode glass slide (cat#48300-026, Avantor). The PDMS layer is a microfluidic chip comprising an inlet accommodating multiple solution ports, a microchannel network, and an outlet. To prevent debris in biological fluids from potentially blocking the microchannels, coarse filters were imprinted prior to the microchannels. The device has a single outlet port for washing buffer waste removal, sample cycle recycling, and collection. Inlet and outlet ports were punched out using a 20-gauge blunt tip needle (cat#90120050D, CML Supply). The PDMS chip was manually aligned to the electrode array under stereo microscopy after the plasma treatment step (Reactive Ion Etch Series 800 Plasma System, MICRO-RIE).

### 4.2   Electrode Design and Modeling

We adopted an electrode design and modeling approach, building upon previously established methodologies[23], with key modifications to enhance throughput while preserving electroporation performance. Our focus was to develop a new electrode arrangement that enhances sample processing throughput while maintaining per-chamber, vortex-assisted electroporation performance comparable to our previous low-throughput design. To enable uniform voltage distribution and synchronized operation across all chambers, the chamber electrodes were connected in a series-parallel configuration, balancing the electrical load while preserving efficient electroporation conditions. We evaluated the patterned Au electrodes by converting them into equivalent circuit diagrams (Figure S3). Each segment of the patterned Au was approximated as a rectangular section with a uniform cross-section and material composition, and the segment resistance, $R$, was computed from resistivity, $\rho$, as:

$$R = \rho \frac{l}{A} = \rho \frac{l}{wh}$$

Here, $\rho$ is the resistivity of Au (2.44 × $10^{-8}$ Ω·m for our calculations), l is the segment length, and $A$ is the cross-sectional area defined by the width, $w$, and height, $h$. The electrode thickness, $t$, was fixed at 300 nm to ensure that stable plasma bonding between the PDMS and the glass slide could occur. Each chamber used five pairs of interdigitated electrodes (450 µm length and 20 µm width) connected in opposite polarity via a common bus line (Figure S3b). The electrical resistance of the per-chamber electrode network was estimated to be 401.6 Ω using COMSOL Multiphysics simulations with DPBS as the cell suspending medium. The final electrode schematic, along with resistance estimates, was translated onto SPICE models to simulate the voltage drop across each electrode (Figure S3b).

### 4.3 Cell Culture

Immortalized cell lines (MCF-7 and HEK 293 cells) were purchased from ATCC and maintained in Dulbecco's Modified Eagle Medium (DMEM) (cat#11995065, Thermo Fisher) supplemented with 10% (v/v) heat-inactivated fetal bovine serum (HI-FBS) (cat#16140071, Thermo Fisher) and 1% (v/v) penicillin-streptomycin (cat#15140122, Thermo Fisher). Cells were passaged at 70% confluency.

HMF (cat#7630, ScienCell) were maintained following manufacturer protocol at 37 °C in a 5% $CO_2$ humidified incubator. Cells were passaged at 90% confluency. PD was calculated by taking the base-2 log of the ratio between the total cells collected before subculture and total cells split into a new vessel. Starting from a PD of 0 at passage 1, the PD of low- and high-passage cells were between 7-14 and 14-30, respectively. Unless otherwise stated, HMF cells with PD < 14 were utilized for all plasmid and mRNA electroporation experiments.

All cells used in our experiments were subcultured using 0.05%(v/v) of Trypsin-EDTA (cat#25200056, Thermo Fisher) diluted in PBS following standard protocol.

### 4.4 Electroporation Buffer Preparation

DPBS (cat#14190144, Thermo Fisher) and Opti-MEM™ (cat#51985034, Thermo Fisher) were utilized as a base electroporation buffer (EP) for this study. Plasmids and mRNA were directly diluted into the base EP buffer. For experiments that added DMSO to the electroporation buffer, 1% (v/v) dimethyl sulfoxide (DMSO) (cat#45001-118, Corning) was added to each electroporation buffer directly prior to electroporation to minimize base pairing interaction with the genetic cargo.

### 4.5 Solution Exchange System

The positive-pressure pneumatic flow control system described in our previous works was utilized for solution exchange in this study[22–24]. To interface this system with standard labware, we 3D-printed a tube holder that couples the pneumatic line to standard 50mL Falcon tubes (cat#21008-178, Avantor), enabling the controlled injection of solutions into the microfluidic devices. To assess cell-trapping efficiency, we connected two solution ports—one for the DPBS wash buffer and the other for the cell suspension—to the device's inlet. For gene delivery experiments, three solution ports—DPBS with 1% (v/v) DMSO wash buffer, the customized electroporation buffer, and the cell suspension (cells resuspended in growth media)—were connected to the inlet of the device.

### 4.6 Trapping Efficiency Experimental Procedure

MCF-7 cells were trypsinized and pelleted by centrifugation at 1100 rpm for 5 min and diluted in fresh media to a final concentration of $5\times10^3$ cells/mL. Cells were then stained with 1 µM Calcein AM (cat#564061, BD Pharmingen) for 10 min prior to cell trapping efficiency experiments. 200 µL of this cell suspension was diluted in 40 mL of media for each syringe test ($1×10^3$ cells, 25 cells/mL) in 50 mL syringes (cat#13-689-8, Fisher Scientific). Syringes were loaded so that the total volume processed would not exceed 80% of the syringe capacity (i.e., for the 50 mL syringe, 200 µL of the suspension were added to 39.8 mL of media). For the cell trapping syringe sequence, formation of vortex trapping flow rates was first primed at 40 psi from the DPBS vial for 30 s. The syringe pump (cat#70-3007, Harvard Apparatus) then infused the cell suspension at 4 mL/min for 50 s (total flow rate through device ~ 5 mL/min, $Re$ = 136), followed by a withdrawal step at 0.5 mL/min for 30 s. This withdrawal step is necessary to ensure that only the vortex-trapped cells are collected during the collection procedure. After processing, cells were collected into 96-well plates (cat#07-000-162, Fisher Scientific) and imaged within 2 hrs. Cell diameters were measured from fluorescence images using an intensity-based image-recognition analysis, as described previously[24].

### 4.7 Membrane-Impermeable Molecule Electroporation Experimental Procedure

MCF-7 cells were trypsinized and pelleted by centrifugation at 1100 rpm for 5 min and resuspended in fresh media to a final concentration of $1×10^3$ cells/mL. Cells were then stained with the nucleus dye NucBlue™ Live ReadyProbes™ according to manufacturer protocol to label viable cells pre-electroporation (cat# R37605, Thermo Fisher). 4 mL of this suspension was loaded into a 5 mL syringe. After the cell trapping syringe sequence, the solution port was switched from the DPBS wash vial to the electroporation buffer vial. All membrane-impermeable molecule delivery experiments were conducted using DPBS as the base electroporation buffer. For YOYO-1 delivery, electrical parameters were 10 pulses of 1 ms AC square waves at 10 kHz, with 1 s inter-pulse interval. Upon completion, cells were collected by reducing the pressure in the solution vials from 40 psi to 30 psi for 10 s, then transferred into 96-well plates at 5 psi for an additional 10 s. Immediately after collection, 100 µL of pre-warmed DPBS containing 2 µM Calcein Red AM (final well concentration ~1µM) was added to each well, and the 96-well plate was placed in a humidified incubator at 37 °C with 5% $CO_2$. At 20 min post-collection, the solution was replaced with 100 µL of pre-warmed DPBS, and wells were imaged using fluorescence microscopy.

### 4.8 Plasmid Preparation and Extraction

The plasmid pZsGreen1-C1 (hereafter referred to as ZsGreen) was obtained from Takara Bio (cat#632447). The plasmid pcDNA-3xHA-hTERT was obtained from Addgene (ID: 51637). The plasmid pcDNA3.1(+) eGFP was obtained from Addgene (ID: 129020). Plasmids were transformed into *E. coli* DH5α cells (cat#18265-017, Thermo Fisher) for amplification, unless stated otherwise. Transformed cells were cultured in Luria-Bertani broth (cat#12795027, Thermo Fisher) prepared according to the manufacturer's protocol. Glycerol stocks were prepared by mixing the inoculated broth 1:1 with 50% (v/v) sterile glycerol (cat#M153-100ML, Pantek Technologies) diluted in autoclaved water, and stored at -80°C.

Glycerol stocks were used to inoculate 5mL overnight starter culture at 37 °C with constant agitation (240 rpm). The starter culture was then expanded to a maxi-scale (>100 mL) culture grown for 12 hr at 37 °C with reduced agitation (160 rpm) to minimize foaming. Plasmids were extracted using the Plasmid Plus Maxi Prep (cat#12963, QIAGEN) by following the manufacturer's protocol. Plasmid concentration and purity were assessed by Nanodrop mode on a spectrophotometer (Take3 Micro-Volume Plate, BioTek Cytation 5).

For quantitative comparison, microscale electroporation transfection efficiencies were normalized to those obtained using conventional transfection agents. Conventional agents were Lipofectamine™ 3000 (cat#L3000001, Thermo Fisher) for plasmid transfection, and Lipofectamine™ MessengerMAX™ (cat#LMRNA001, Thermo Fisher) for mRNA transfection. All transfections followed the manufacturers' protocols. Fluorescence thresholds were set to the maximum signal emitted from non-electroporated cells.

### 4.9    In vitro mRNA Synthesis

The plasmid pcDNA3.1(+) eGFP (hereafter referred to as eGFP) was obtained from Addgene (ID: 129020). Glycerol stocks were used to inoculate 5mL overnight starter cultures at 37 °C with constant agitation (240 rpm). The plasmid was then extracted using a spin miniprep kit (cat#27104, QIAGEN) following the manufacturer's protocol and linearized by a single-site restriction enzyme digestion (BbsI-HF, cat#R3539S or PmeI, cat#R0560S, New England Biolabs) according to the manufacturer's instructions. In vitro transcription (IVT) was carried out using T7 RNA polymerase (cat# M0251S, New England Biolabs) with either CleanCap Reagent AG (cat#N-7113, TriLink Biotechnologies) or Anti-Reverse Cap Analog (ARCA) (cat# S1411S, New England Biolabs), substituting uridine entirely with N1-Methylpseudouridine-5'-Triphosphate (N1MePsU) (cat#N-1081, TriLink Biotechnologies). The correct 5'-AG-3' initiating sequence immediately following the T7 promoter site for CleanCap Reagent AG mRNA synthesis was introduced via site-directed mutagenesis using the Q5 Hot Start High-Fidelity DNA Polymerase Kit (cat#M0493S, New England Biolabs), with the following primers: forward:"- ACT CAC TAT AAG GAG ACC CAA G -" and reverse:"- CGT ATT AAT TTC GAT AAG CCA G -"(Integrated DNA Technologies). The mutation in the template was confirmed by Sanger sequencing. ARCA-capped mRNA was synthesized from the unmutated template. A Poly-(A) tail of approximately 100 nt was added using *E. coli* Poly(A) Polymerase (cat#M0276L, New England Biolabs) to the mRNA product. After each step, RNA products were purified using an RNA Cleanup Kit (cat#T2050S, New England Biolabs). RNA concentration and purity were assessed using a Nanodrop spectrophotometer. Commercially available eGFP mRNA with 5-methoxyuridine (5moU) uridine substitution (cat#L-7201, TriLink Biotechnologies) was purchased to compare the translation efficiency of in vitro transcribed mRNA products.

### 4.10    Electroporation-mediated Transfection Procedure

HEK293 and HMF were trypsinized, pelleted by centrifugation at 1100 rpm for 5 min, and resuspended in fresh media to a final concentration of $1 \times 10^3$ cells/mL. 4 mL of this suspension was loaded into a 5 mL syringe (cat# 75846-756, Avantor Science). Following the syringe sequence described in previous sections, cells were trapped in microvortices for an additional 2 min, then the active solution port was switched from the wash buffer to the electroporation buffer. Electroporation pulses were applied while maintaining an infusion pressure of 40 psi in the electroporation buffer. Unless otherwise stated, the electrical parameters for gene delivery were 20 pulses of 1 ms AC square waves at 10 kHz, with 1 s inter-pulse interval. Upon completion, cells were collected by reducing the vial pressure from 40 psi to 30 psi for 10 s, then transferring them into 96-well plates at 5 psi for an additional 10 s. Immediately after collection, 100 µL of pre-warmed medium was added to each well, and the 96-well plate was placed in a humidified incubator at 37 °C with 5% $CO_2$. At 1 hr post-collection, the medium was replaced with 100 µL of pre-warmed medium to completely remove residual electroporation buffer. Transfection outcomes were quantified by imaging at 24 hr (mRNA) and 48 hr (plasmid) post-electroporation to enumerate reporter-positive cells (e.g., GFP or HA), using direct reporter fluorescence for GFP and immunofluorescence for HA.

### 4.11    Immunofluorescence

All immunofluorescence experiments were carried out in 96-well plates unless stated otherwise. Cells were fixed with 4% paraformaldehyde (cat#15710, Electron Microscopy Sciences) for 20 min and permeabilized with 0.1% Triton X-100 (cat#T8787, Sigma Aldrich) in DPBS for 2 min. Primary antibodies against HA (rabbit anti-HA C29F4 cat#3724S, Cell Signaling Technology, 1:200 dilution) or Ki67 (mouse anti-Ki67 8D5 cat#9449S, Cell Signaling Technology, 1:500 dilution) were diluted in 10% normal goat serum (cat#50062Z, Thermo Fisher) and incubated for 1 hr at room temperature in a humidified chamber. Subsequently, cells were incubated with Alexa Fluor-conjugated secondary antibodies (anti-rabbit cat#4414S, Cell Signaling Technology, 1:3000 dilution) (anti-mouse cat#A-11001, Thermo Fisher. 1:1000 dilution) and DAPI nuclei counterstain (0.2 µg/well; cat#D3571, Thermo Fisher), both diluted in 10% normal goat serum, for 1 hour at room temperature in a humidified chamber. Between each step, each well was washed three times with PBS-Tween (0.05% v/v; cat#85113, Thermo Fisher).

## 5 Acknowledgments

This work is partially supported by Susan G. Komen Career Catalyst Research Grant (CCR19609203) and NIH R21CA229024. The authors would like to thank Dr. Jung Soo Suk at University of Maryland for kindly gifting us the plasmid pZsGreen1-C1, Huy Vo and the Whiting School of Engineering Whitaker Microfabrication Lab for support with photolithography and device fabrication, Dr. Jeff Coller and Dr. Zachary Mandell from the Johns Hopkins University RNA Innovation Center for guidance with mRNA in vitro transcription, and the Genetic Resources Core Facility (GRCF) at the Johns Hopkins Medical Institute for sequencing. We would like to acknowledge editing support by Anne N. Connor, provided by the Office of the Vice Provost for Research at Johns Hopkins University. ChatGPT, a language model developed by OpenAI, was utilized to improve clarity of the manuscript's wording.

**Data availability**

The data supporting the findings of this study are available from the corresponding author upon reasonable request.

**Conflicts of interest**

SCH has a financial interest in Vortex Biosciences, Inc. and receives royalties related to intellectual property utilized in this study. HWS declares no potential conflict of interest.

**Supplementary Materials**

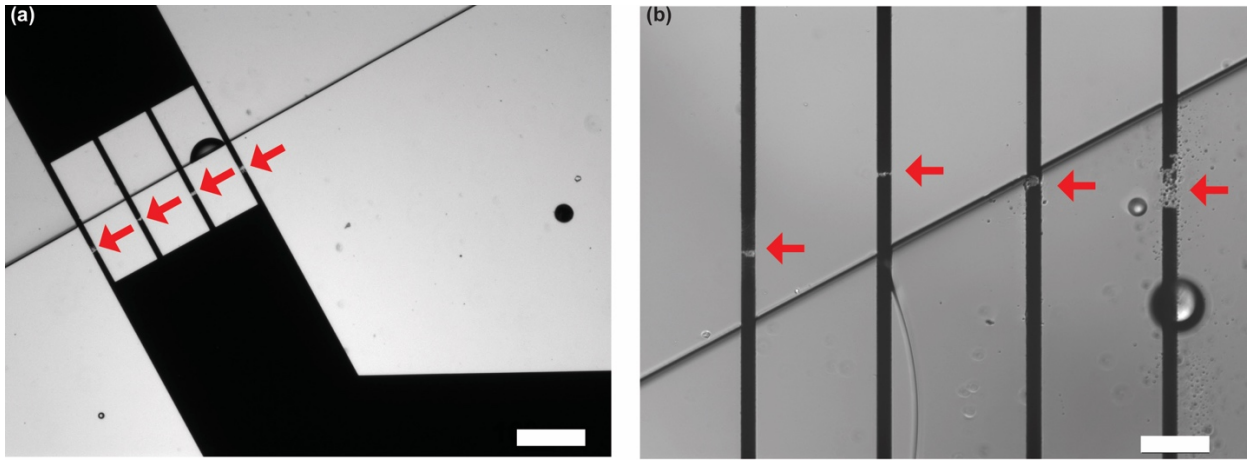

**Figure S1: Electrode degradation under high-voltage electroporation conditions.** (a) Brightfield micrograph showing electrode erosion and void formation (red arrows) along common bus lines within thin electrode regions following high-voltage operation. Degradation is most pronounced in regions of elevated electrical resistance, consistent with localized current crowding and electrochemical damage. Scale bar = 250 µm. (b) Higher-magnification view of representative erosion and void formation sites corresponding to the regions indicated in (a). Scale bar = 100 µm.

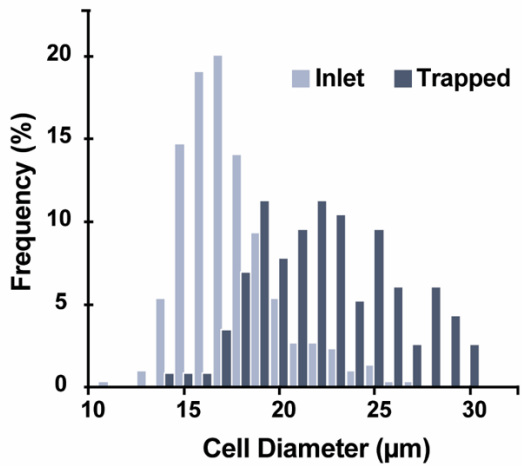

**Figure S2 Diameter distribution of MCF-7 cells at the device inlet and after vortex trapping.** Overlaid histograms show the cell size distribution at the inlet (light gray) and for vortex-trapped cells (dark gray), revealing selective enrichment of larger cells. Mean cell diameters increased from 16.9 ± 2.5 µm at the inlet to 22.1 ± 3.6 µm after trapping.

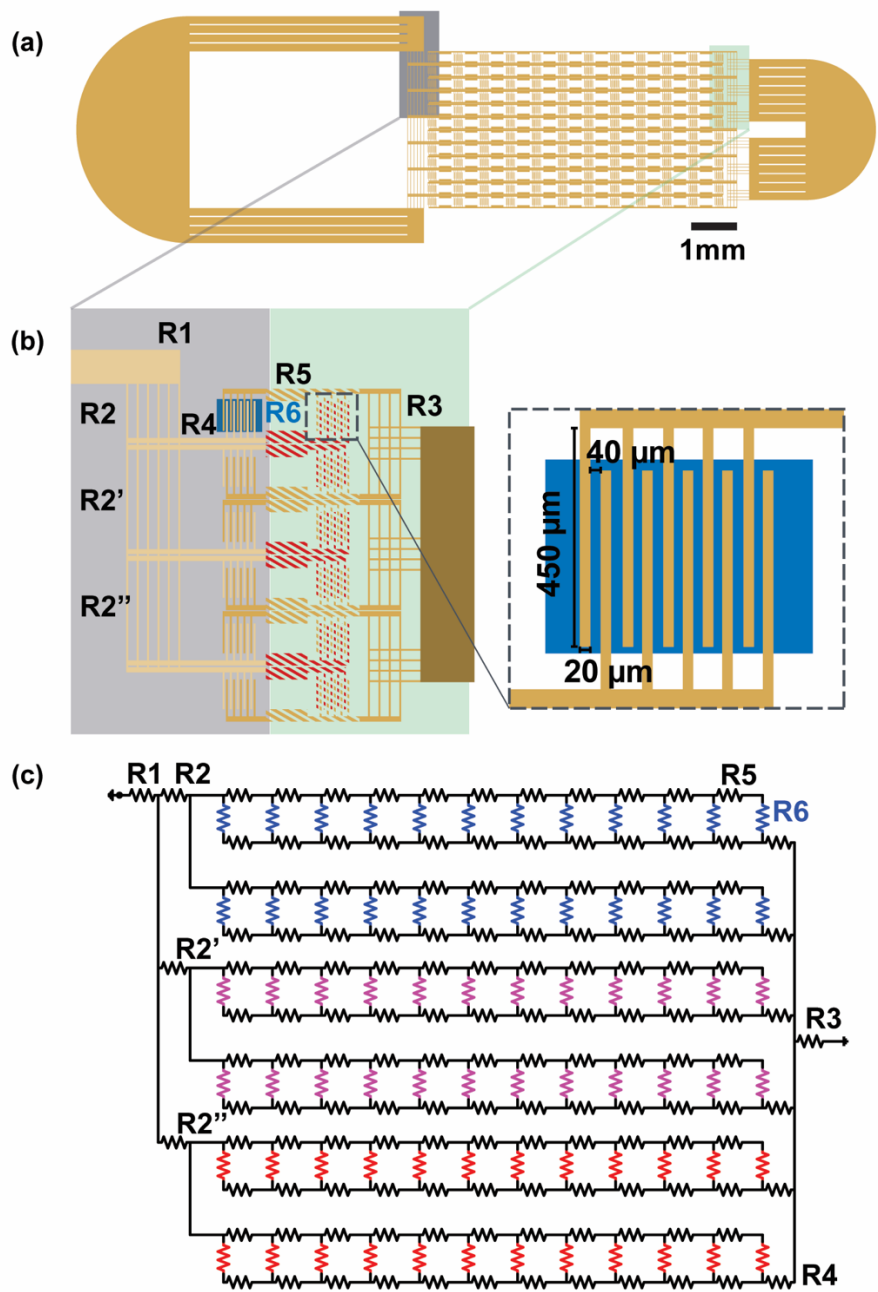

**Figure S.1.13: Device architecture and electrical modeling schematic.** (a) CAD rendering of the patterned Au electrode layout used for vortex-assisted electroporation, showing the overall electrode geometry and routing pathways. The dashed box highlights the electroporation chamber array. Scale bar = 1 mm. (b) Expanded view of the electrode layout and routing architecture. Labeled resistor elements (R1–R6) correspond to distinct electrode and routing regions used for circuit modeling. Geometric dimensions and resistance values for each region are summarized in **Table S2.1.1.** The dashed box shows a magnified view of a single electroporation chamber comprising five pairs of interdigitated electrodes, with key

geometric dimensions indicated. (c) Equivalent electrical resistance network used for SPICE modeling of the electrode array. Individual resistors represent geometrically distinct electrode and routing segments, with color coding corresponding to different resistance regions. Only the upper half of the network is shown, as device symmetry results in identical resistive behavior in both halves.

| Name of Resistor | Component | Inlet/Outlet | Length (µm) | Width (µm) | CS Area (µm2) | R (Ω) | # in Parallel | Parallel Res. (Ω) | Total Resistance |
|---|---|---|---|---|---|---|---|---|---|
| R1 | E1 | Inlet | 15440 | 500 | 150 | 2.51 | 4 | 0.63 | **0.63** |
| R2 | E2 | Inlet | 760 | 20 | 6 | 3.09 | 6 | 0.52 | **0.92** |
| R2' | E3 | Inlet | 2484 | 20 | 6 | 10.10 | 6 | 1.68 | **2.09** |
| R2'' | E4 | Inlet | 4280 | 20 | 6 | 17.41 | 6 | 2.90 | **3.31** |
|  | E1 | In-Between | 800 | 80 | 24 | 0.81 | 2 | 0.41 |  |
| R6 | Chamber | In-Between |  |  | COMSOL Simulation |  |  |  | **401.77** |
| R5 | E1 | Outlet | 920 | 80 | 24 | 0.94 | 1 | 0.94 |  |
|  | E2 | Outlet | 800 | 180 | 54 | 0.36 | 1 | 0.36 | **1.30** |
| R4 | E1 | Inlet | 800 | 80 | 24 | 0.81 | 1 | 0.81 |  |
|  | E2 | Inlet | 451 | 20 | 6 | 1.83 | 4 | 0.46 | **1.27** |
| R3 | E1 | Outlet | 800 | 20 | 6 | 3.25 | 12 | 0.27 |  |
|  | E2 | Outlet | 3080 | 500 | 150 | 0.50 | 12 | 0.04 | **0.31** |

| Constants | Resistivity | 0.0244 | [Ω*µm] |
|---|---|---|---|
|  | Thickness | 0.3 | µm |

**Table S1: Electrical resistance parameters for SPICE modeling of the vortex-assisted electroporation device**. Geometric dimensions of electrode and routing regions corresponding to resistors R1–R6 in **Figure S2.1.1** were extracted from CAD designs and converted to electrical resistance values using the resistivity of gold (0.0244 Ω·µm) and a deposited electrode thickness of 0.3 µm. Resistance calculations account for segment length, width, cross-sectional area, and parallel circuit configurations. Individual resistances were combined according to circuit topology to determine the effective resistance of each device region. The electroporation chamber resistance (R6) was obtained from COMSOL finite-element simulations. These parameters were used for SPICE-based modeling of voltage distribution and electric field uniformity.

E1–E4 denote electrode regions; CS area, cross-sectional area; # in parallel, number of parallel resistive elements; Parallel Res., equivalent parallel resistance.

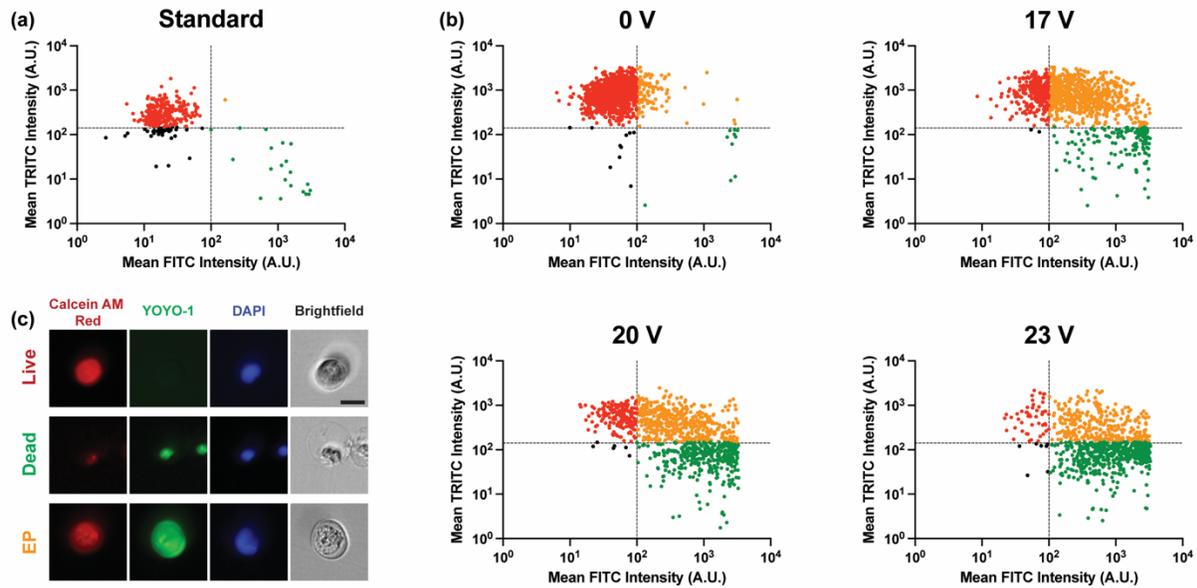

**Figure S4. Gating strategy for electroporation efficiency determination.** (a) Standard control used to establish fluorescence intensity thresholds for cell viability and electroporation status. The viability threshold (Calcein Red-AM, y-axis) was set at the maximum fluorescence intensity of dead cells to minimize false-positive classification. The electroporation threshold (YOYO-1, x-axis) was defined such that fewer than 1% of viable control cells exceeded this value. (b) Application of the standardized gating thresholds to electroporated cell populations across applied voltages (0–23 V). Color-coded subpopulations correspond to viable non-electroporated cells (red), viable electroporated cells (orange), and dead or lysed cells (green) attributed to electroporation-induced membrane damage. Events falling below both viability and electroporation thresholds (black) were excluded from efficiency calculations. Voltage-dependent shifts in population distributions indicate the emergence of optimal electroporation conditions at intermediate voltages. (c) Representative fluorescence and brightfield images of viable, electroporated, and dead cells, showing Calcein Red-AM, YOYO-1, DAPI, and corresponding brightfield channels. Scale bar = 20 µm.

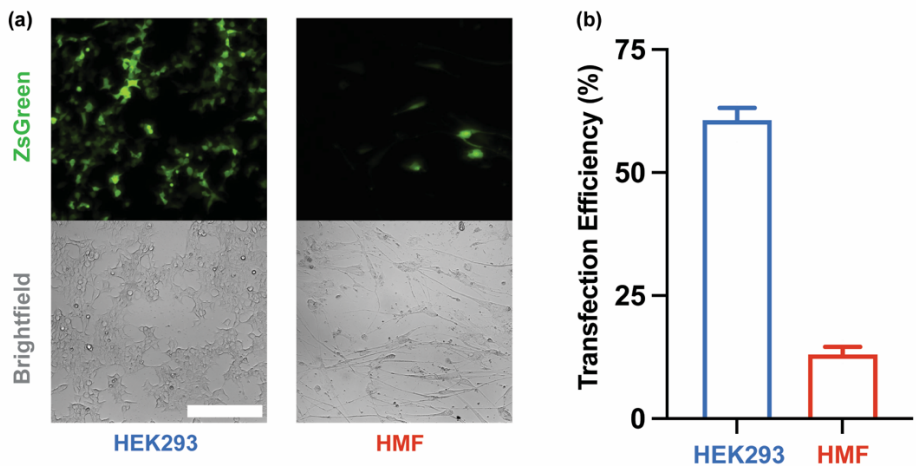

**Figure S5: Comparison of conventional transfection efficiency in immortalized and primary cells.** (a) Representative fluorescence (ZsGreen) and corresponding brightfield microscopy images of HEK293 (immortalized) and human mammary fibroblast (HMF, primary) cells 24 h post-transfection using a conventional chemical transfection reagent. Scale bar = 100 µm. (b) Quantification of transfection efficiency at 24 h post-transfection, calculated as the fraction of ZsGreen-positive cells relative to the total DAPI-positive cell population. Data are presented as mean ± SEM (n = 3; >200 cells per replicate).

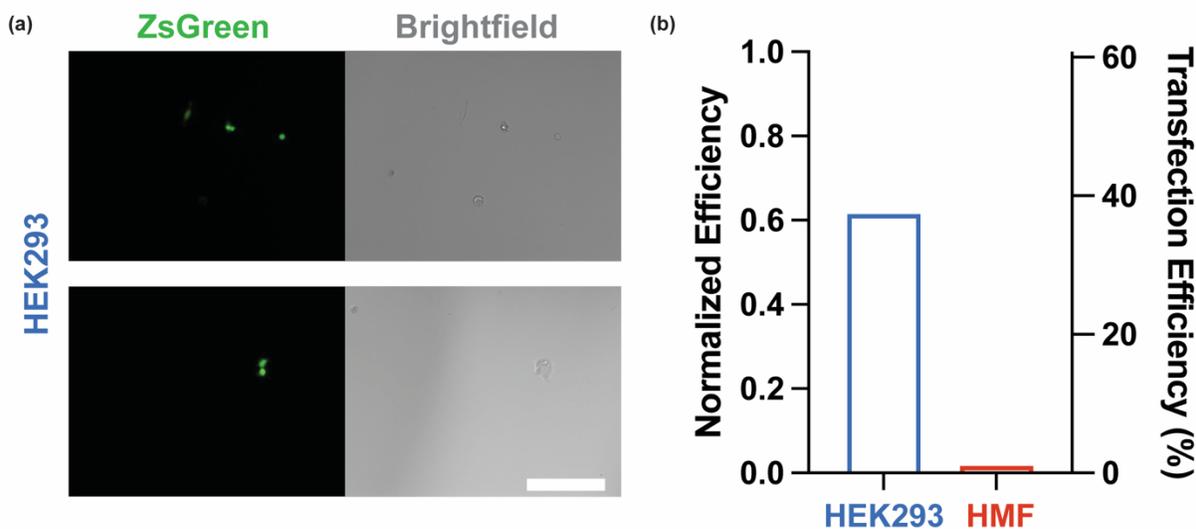

**Figure S6: Comparative vortex-assisted electroporation efficiency in immortalized and primary cells.** (a) Representative fluorescence (ZsGreen) and corresponding brightfield microscopy images of HEK293 (immortalized) cells following device-based electroporation in DPBS base buffer without DMSO (50 µg/mL plasmid; 20 V input, 10 kHz frequency, 20 pulses). Images were acquired 24 h post-electroporation. Lower apparent cell density in

HEK293 samples reflects reduced trapping efficiency associated with smaller cell size. HFM (primary) cells are not shown. Scale bar = 100 μm. (b) Quantitative comparison of electroporation efficiency between HEK293 and HMF cells under identical operating conditions. The left y-axis shows electroporation efficiency normalized to Lipofectamine-mediated transfection efficiency measured for each cell type (Figure S5), enabling cell-type-independent comparison of delivery performance. The right y-axis shows the corresponding observed electroporation transfection efficiency.

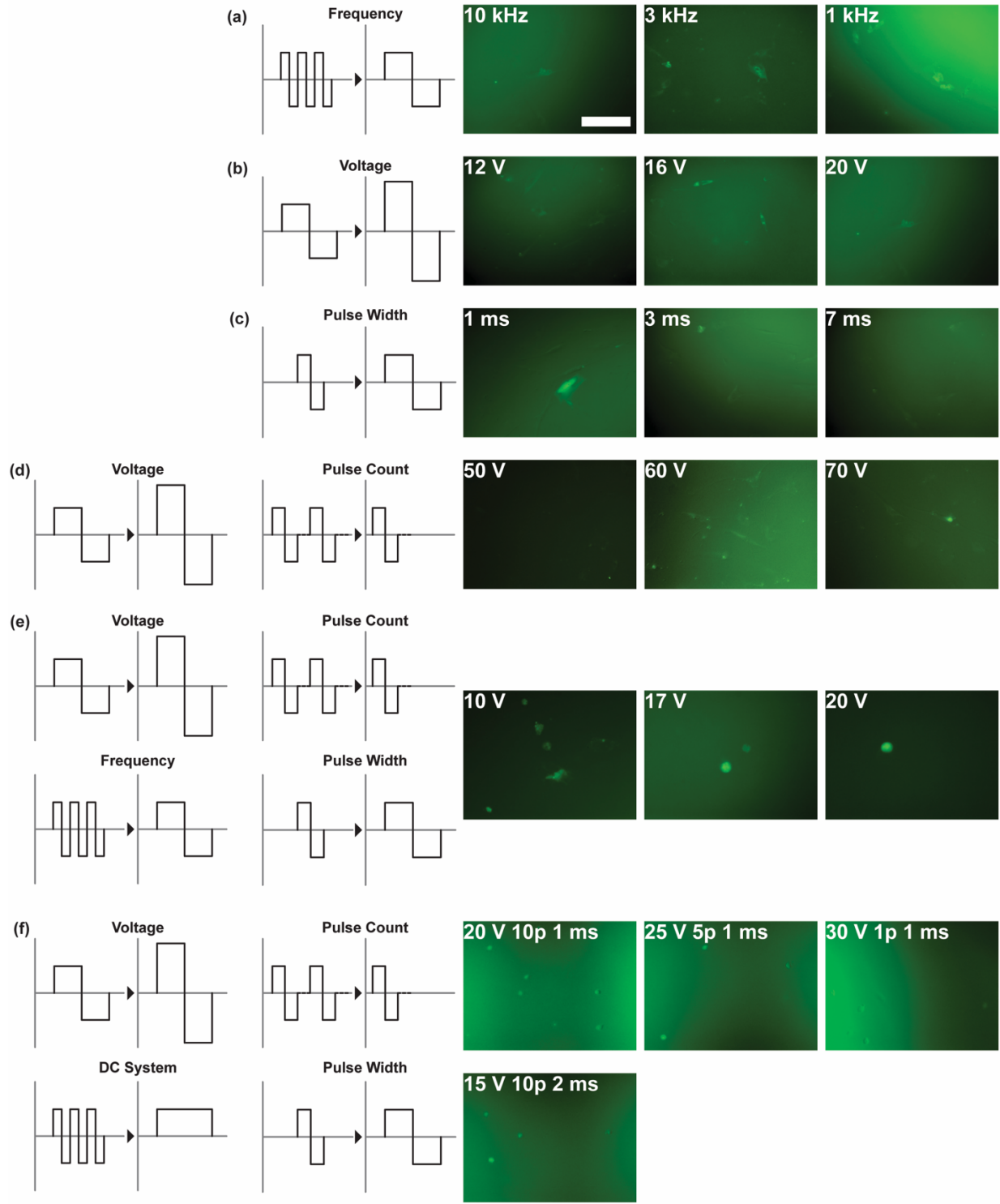

**Figure S7: Systematic electrical parameter optimization demonstrating insufficient transfection in primary cells.** Schematic waveforms (left) illustrate the electrical parameters varied for each condition. Unless otherwise noted, base electroporation parameters consisted of 20 pulses of 1 ms AC square waves at 20 V and 10 kHz, with a 1 s

inter-pulse interval. Fluorescence microscopy images (right) show ZsGreen expression in HMF cells acquired 72 h post-electroporation. Nomenclature: V, voltage; kHz, kilohertz; p, pulse count. (a–c) Single-parameter optimization experiments showing electroporation outcomes with varying (a) pulse frequency (10, 3, and 1 kHz), (b) voltage amplitude (12, 16, and 20 V), and (c) pulse width (1, 3, and 7 ms). (d) Dual-parameter optimization combining elevated voltage amplitudes (50–70 V) with reduced pulse counts (2 pulses) to mitigate electrode erosion. (e) Four-parameter multiplexed optimization integrating voltage, pulse count, frequency, and pulse width. Representative conditions (10, 17, and 20 V) exhibit persistently low fluorescence intensity despite increased waveform complexity. (f) Multiplexed optimization using a DC electroporation system. The top row shows pulsed conditions (20 V, 10 pulses, 1 ms; 25 V, 5 pulses, 1 ms; 30 V, 1 pulse, 1 ms), while the bottom row shows DC conditions (15 V, 100 pulses, 2 ms).

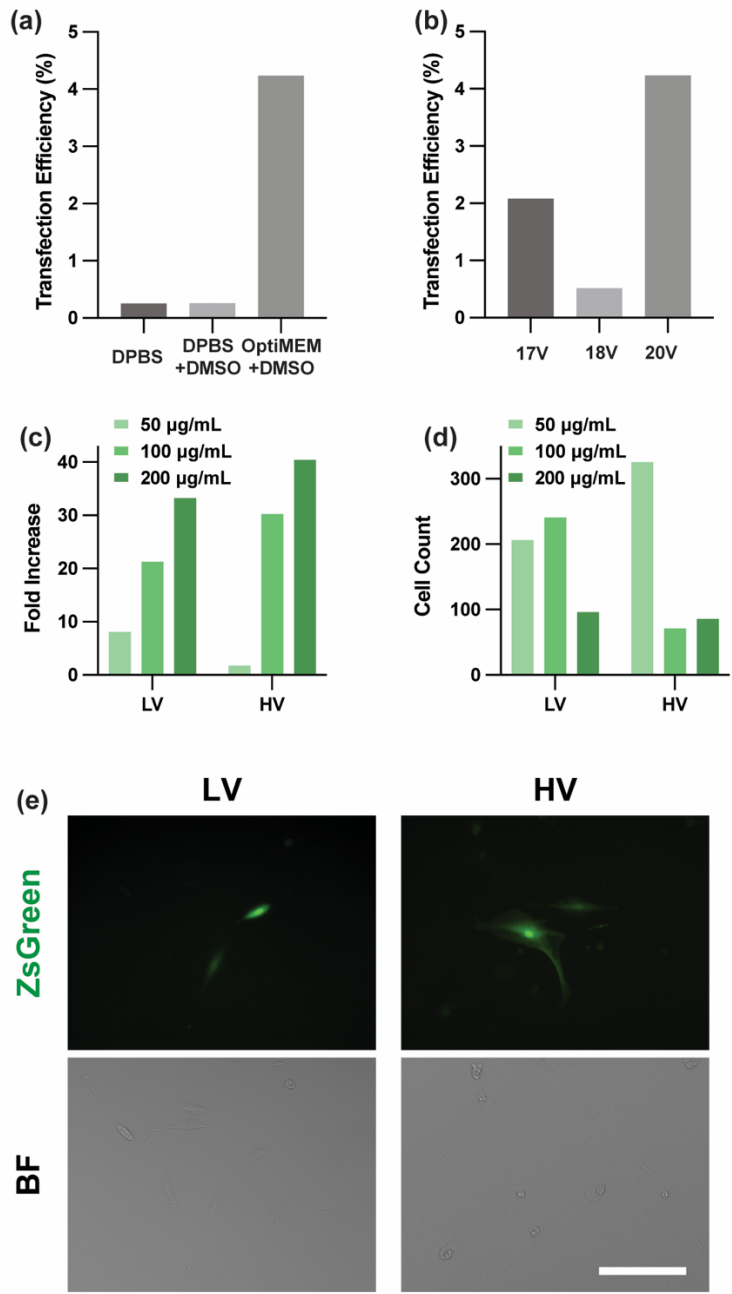

**Figure S8 Buffer- and voltage-dependent modulation of plasmid electroporation in primary cells.** (a) Absolute ZsGreen plasmid transfection efficiency at 20 V using 50 µg/mL plasmid, comparing DPBS, DPBS + DMSO, and Opti-MEM + DMSO buffer formulations. (b) ZsGreen plasmid transfection efficiency as a function of applied voltage (17, 18, and 20 V) using Opti-MEM + DMSO with a fixed plasmid concentration of 50 µg/mL. (c) Fold increase in transfection efficiency relative to the DPBS + DMSO baseline for low-voltage (LV) and high-voltage (HV) regimes across plasmid concentrations (50, 100, and 200 µg/mL). (d) Collected cell counts corresponding to the LV and HV conditions shown in (c), illustrating reduced

recovery at higher voltages. (e) Representative fluorescence (ZsGreen) and corresponding brightfield (BF) microscopy images of HMF cells electroporated under LV and HV conditions using 200 µg/mL plasmid. Increased incidence of burst cells is observed under HV conditions. Scale bar = 100 µm.

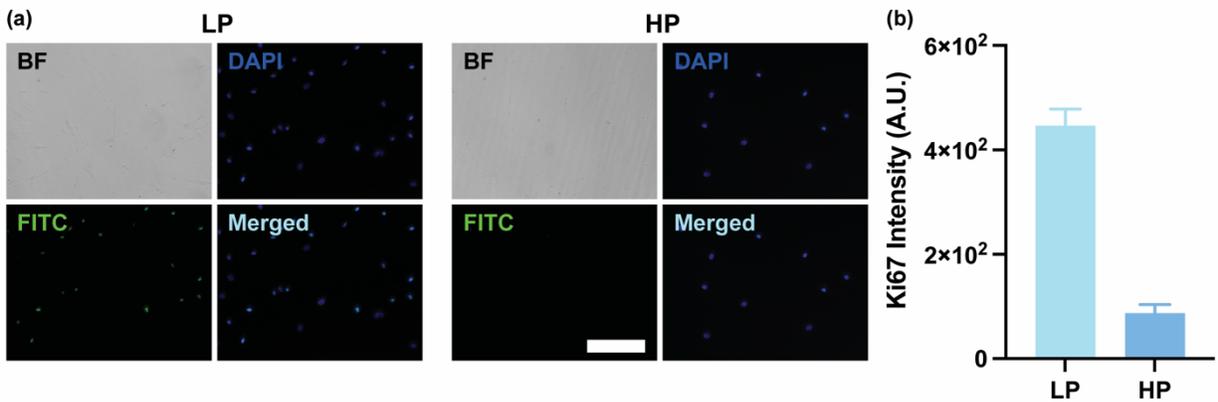

**Figure S9: Ki67 expression in low- and high-passage HMF cells.** (a) Representative brightfield (BF) and fluorescence microscopy images of Ki67 immunostaining in low-passage (LP, left) and high-passage (HP, right) human mammary fibroblast (HMF) cells. Images show DAPI nuclear staining (blue), Ki67 immunofluorescence (FITC, green), and merged channels. Scale bar = 100 µm. (b) Quantitative analysis of Ki67 fluorescence intensity per cell, demonstrating significantly higher proliferative activity in LP compared to HP HMF cells. Error bars represent mean ± SEM (>200 cells per condition).

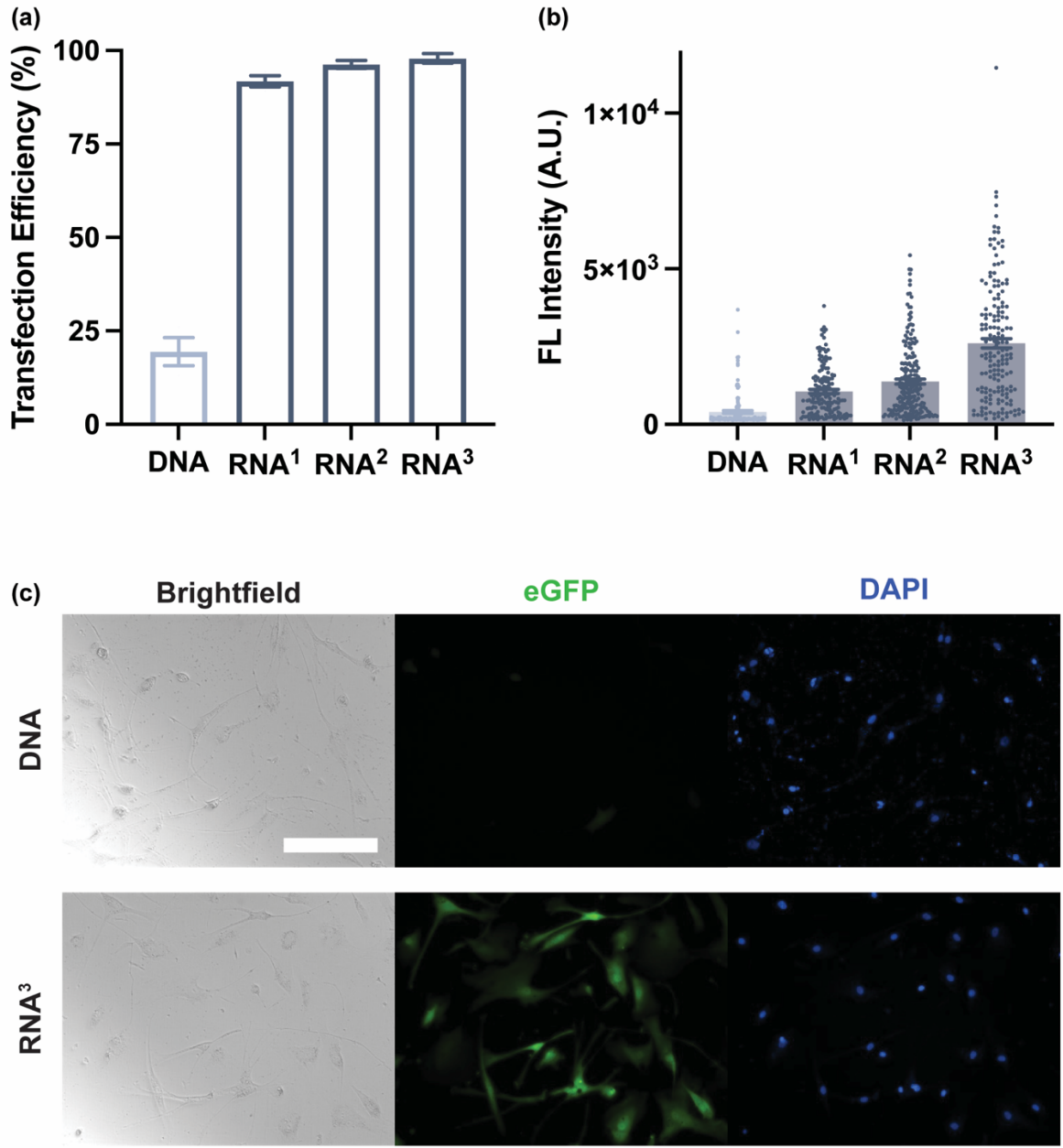

**Figure S10: Comparison of plasmid DNA and mRNA transfection using conventional reagents in primary HMF cells.** (a) Transfection efficiency comparison between eGFP plasmid DNA and three eGFP mRNA variants: RNA$^1$ (ARCA cap, N1-methylpseudouridine), RNA$^2$ (CleanCap AG, 5-methoxyuridine), and RNA$^3$ (CleanCap AG, N1-methylpseudouridine). mRNA nomenclature denotes cap analog and uridine modification, respectively. Error bars represent mean ± SEM (n = 3). (b) Single-cell fluorescence intensity analysis showing differences in expression levels among cargo types despite comparable transfection efficiencies. RNA$^3$ exhibited the highest eGFP expression in HMF cells and was selected for subsequent device-based mRNA electroporation experiments. Error bars represent mean ±

SEM (n > 150 cells). (c) Representative brightfield and fluorescence microscopy images of HMF cells transfected with eGFP plasmid DNA (DNA) or eGFP mRNA (RNA$^3$) using cargo-specific conventional transfection reagents. Images show brightfield (left), eGFP fluorescence (center), and DAPI nuclear staining (right). Scale bar = 100 µm.